\documentclass[showpacs,amsmath,amssymb,prl,superscriptaddress,twocolumn]{revtex4}
\usepackage[T1]{fontenc}
\usepackage{amssymb,amsmath}
\usepackage{times}
\usepackage{graphicx}
\usepackage{wasysym}
\usepackage{subfigure}
\usepackage[sort&compress]{natbib}
\usepackage{bm}
\usepackage{overpic}
\usepackage{color}

\begin{document}

\title{Theory of pump-probe experiments of metallic metamaterials coupled to the gain medium}

\author{Zhixiang Huang}
\affiliation{Ames Laboratory and Dept.~of Phys.~and Astronomy,
             Iowa State University, Ames, Iowa 50011, U.S.A.}
\affiliation{Key Laboratory of Intelligent Computing and Signal Processing,
             Anhui University, Hefei 230039, China}

\author{Th.~Koschny}
\affiliation{Ames Laboratory and Dept.~of Phys.~and Astronomy,
             Iowa State University, Ames, Iowa 50011, U.S.A.}

\author{C.~M.~Soukoulis}
\affiliation{Ames Laboratory and Dept.~of Phys.~and Astronomy,
             Iowa State University, Ames, Iowa 50011, U.S.A.}
\affiliation{Institute of Electronic Structure \& Laser, FORTH,
             71110 Heraklion, Crete, Greece}

\date{\today}

\begin{abstract}
We establish a new approach for pump-probe simulations of metallic
metamaterials coupled to the gain materials.  It is of vital importance to
understand the mechanism of the coupling of metamaterials with the gain medium.
Using a four-level gain system, we have studied light amplification of arrays
of metallic split-ring resonators (SRRs) with a gain layer underneath. We find
that that the differential transmittance $\Delta T/T$ can be negative for SRRs on the
top of the gain substrate, which is not expected, and $\Delta T/T$ is positive for the
gain substrate alone. These simulations agree with pump-probe experiments and
can help to design new experiments to compensate the losses of metamaterials.
\end{abstract}



\maketitle

The field of metamaterials has been spectacular experiments progress in recent
years \cite{Shalaev2007,Soukoulis2007,Soukoulis2011}.  The mostly metamaterial
are metal-based nano-structure and eventually suffering from the conductor
losses at optical frequencies, which are stills orders of magnitude too large
for the realistic applications. In addition, metamaterial losses become an
increasingly important issue when moving from multiple metal-based metamaterial
layers to the bulk case \cite{Soukoulis2011}.  Thus, the need for reducing or
even compensating of the losses is a key challenge for metamaterial
technologies. One promising way of overcoming the losses is based on
introducing the gain material to the metamaterial. The idea of combination of a
metamaterial with an optical gain material has been investigated by several
theoretical \cite{Fang2010,Wuestner2010,Fang2011,Sivan2009} and experimental
studies \cite{Xiao2010,Plum2009,Tanaka2010,Meinzer2010,Meinzer2011}.
From the experiments point of view, the realistic gain can be
experimentally realized with fluorescent dyes \cite{Xiao2010}, quantum dots
\cite{Plum2009,Tanaka2010} or semiconductor quantum wells
\cite{Meinzer2010,Meinzer2011}.  All these loss-compensation are mainly
attributed to the coupling between metamaterial and the gain medium. Without
sufficient coupling, no loss-compensation can happen, nor can the transmitted
signal be amplified.  Therefore, it is of vital importance to understand the
mechanism of the coupling between metamaterial and the gain medium. In
addition, these ideas can be used in plasmonics to incorporate gain
\cite{Bergman2003,Stockman2008} to obtain new nano-plasmonic lasers
\cite{Oulton2009,Noginov2009}.

In this Letter, we present a systematic theoretical model for pump-probe
experiments of metallic metamaterials coupled with the gain material, described
by a generic four-level atomic system. We describe the dynamical processes in
metamaterials with gain, and increasing the gain changes the metamaterials
properties and we need to have self-consistent calculations
\cite{Fang2010,Wuestner2010,Fang2011} to reach a steady state. The pump-probe
results affecting the time dependence of the population inversion and the
electric field enhancement that increases the effective gain. We observe
differential transmittance signals from the coupled system that are larger than
for the bare gain. Furthermore, we observe a more rapid temporal decay of the
differential transmittance signal for the coupled system compared to the bare
gain. Both effects indicate substantial local-field-enhancement effects, which
increase the effective metamaterial gain beyond the bare gain, leading to a
significant reduction of the metamaterial's losses.

We model the dispersive Lorentz active medium using a generic four-level atomic
system. The population density in each level is given by $N_{i}$ ($i$=0,1,2,3).
The time-dependent Maxwell's equations for isotropic media are given by
$\nabla\times \textbf{E}(r,t) = -\partial \textbf{B}(r,t)/\partial t$ and
$\nabla\times \textbf{H}(r,t) = \partial \textbf{D}(r,t)/\partial t$, where
$\textbf{B}(r,t)=\mu\mu_o \textbf{H}(r,t)$,
$\textbf{D}(r,t)=\varepsilon\varepsilon_0\textbf{E}(r,t)+\textbf{P}(r,t)$ and
$\textbf{P}(r,t)$ is the dispersive electric polarization density that
corresponds to the transitions between two atomic levels, $N_1$ and $N_2$.  The
vectors $\textbf{P}$ introduces gain in Maxwell's equations and its time
evolution can be shown to follow that of a homogeneously broadened Lorentzian
oscillator driven by the coupling between the population inversion and external
electric field \cite{Siegman1968}. Thus, $\textbf{P}$ obeys the equation of
motion
\begin{displaymath}
\frac{\partial^2 \textbf{P}(r,t)}{\partial t^2} +
 \Gamma_a\frac{\partial \textbf{P}(r,t)}{\partial t} +
 \omega_a^2 \textbf{P}(r,t) \ =\
 \sigma_a \Delta N(r,t) \textbf{E}(r,t)
\end{displaymath}
where $\Gamma_a$ stands for the linewidth of the atomic transitions at
$\omega_a$, and accounts for both the nonradiative energy decay rate, as well
as dephasing processes that arise from incoherently driven polarizations. In
the following simulations, this value is equal to $2\pi \times 20 \cdot
10^{12}\,\mathrm{rad/s}$.  $\sigma_a$ is the coupling strength of $\textbf{P}$
to the external electric field and its value is taken to be $10^{-4}\,\mathrm
{C^2/kg}$. The factor $\Delta N(r,t) = N_1(r,t) - N_2(r,t)$ is the population
inversion between level $2$ and level $1$ that drives the polarization
$\textbf{P}$.
\begin{figure}[htbp]
 \centering
 \includegraphics[width=0.45\textwidth]{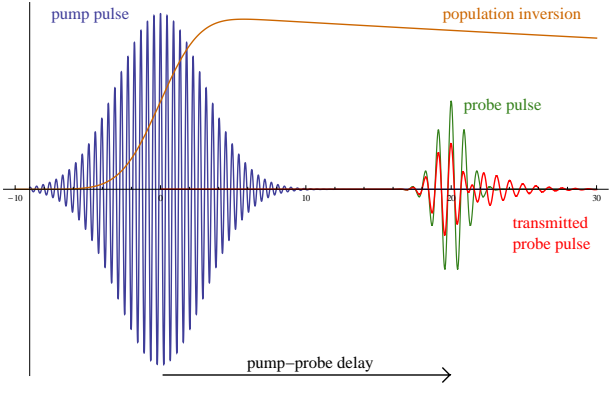}
 \caption{(Color online) Schematic illustration of pump-probe experiments.
 }
 \label{fig1}
\end{figure}
In order to do pump-probe experiments numerically we first pump the gain
material with a short, intense Gaussian pump pulse. After a suitable time delay
we probe the structure with a weak probe pulse (see Fig.~\ref{fig1}).
In our model, an external mechanism pumps electrons from the ground state level
$N_0$ to the third level $N_3$ using a gaussian pumping $P_g(t)$, which is
proportional to the pumping intensity in the experiments. After a short
lifetime $\tau_{32}$ electrons transfer non-radiative into metastable second
level $N_2$.  The second level ($N_2$) and the first level ($N_1$) are called
the upper and lower lasing levels. Electrons can be transferred from the upper
to the lower lasing level by spontaneous and stimulated emission. At last,
electrons transfer quickly and non-radiative from the first level ($N_1$) to
the ground state level ($N_0$).  The lifetimes and energies of the upper and
lower lasing levels are $\tau_{21},\ E_2$ and $\tau_{10},\ E_1$, respectively.
The center frequency of the radiation is $\omega_a=(E_2-E_1)/\hbar$ which is a
controlled variable chosen according to the pump-probe experiments.  The
parameters $\tau_{32}$, $\tau_{21}$, and $\tau_{10}$ are chosen to be
0.05ps, 80ps, and 0.05ps, respectively.
The initial electron density,
$N_0(r,t=0)=5.0\times 10^{23}\,\mathrm {m^{-3}}$, $N_i(r,t=0)=0\,\mathrm
{m^{-3}}$ ($i$=1,2,3).  Thus, the atomic population densities obey the
following  rate equations:
\begin{eqnarray*}
 \frac{\partial N_3(r,t)}{\partial t} &=&
  P_g(t) N_0(r,t) -
  \frac{N_3(r,t)}{\tau_{32}}
  \\
 \frac{\partial N_2(r,t)}{\partial t} &=&
  \frac{N_3(r,t)}{\tau_{32}} +
  \frac{1}{\hbar\omega_a} \textbf{E}(r,t)\cdot\frac{\partial \textbf{P}(r,t)}{\partial t} -
  \frac{N_2(r,t)}{\tau_{12}}
  \\
 \frac{\partial N_1(r,t)}{\partial t} &=&
  \frac{N_2(r,t)}{\tau_{12}} -
  \frac{1}{\hbar\omega_a} \textbf{E}(r,t)\cdot\frac{\partial \textbf{P}(r,t)}{\partial t} -
  \frac{N_1(r,t)}{\tau_{10}}
  \\
 \frac{\partial N_0(r,t)}{\partial t} &=&
  \frac{N_1(r,t)}{\tau_{10}} -
  P_g(t) N_0(r,t)
\end{eqnarray*}
where Gaussian pump $P_g(t)=P_0 \times e^{-(\frac{t-t_p}{\tau_p})^{2}}$, with
$P_0=3 \times 10^{9}\,\mathrm{s^{-1}}$, $t_p=6\,\mathrm{ps}$ \cite{Note1}, and
$\tau_p=0.15\,\mathrm{ps}$.

In order to solve the response of the active materials in the electromagnetic
fields numerically, the FDTD technique is utilized \cite{Taflove1995}, using an
approach similar to the one outlined in \cite{Fang2010b}.

\begin{figure}[htbp]
 \centering
 \includegraphics[width=0.45\textwidth]{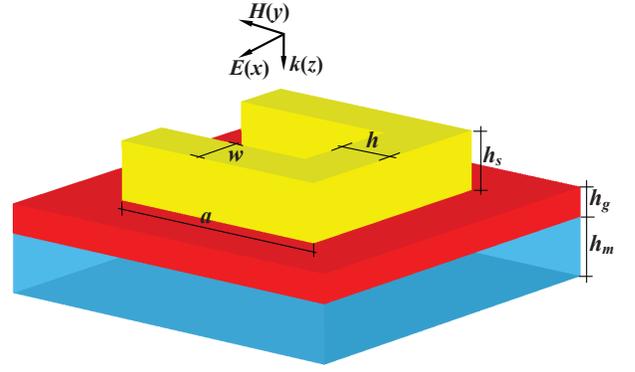}
 \includegraphics[width=0.45\textwidth]{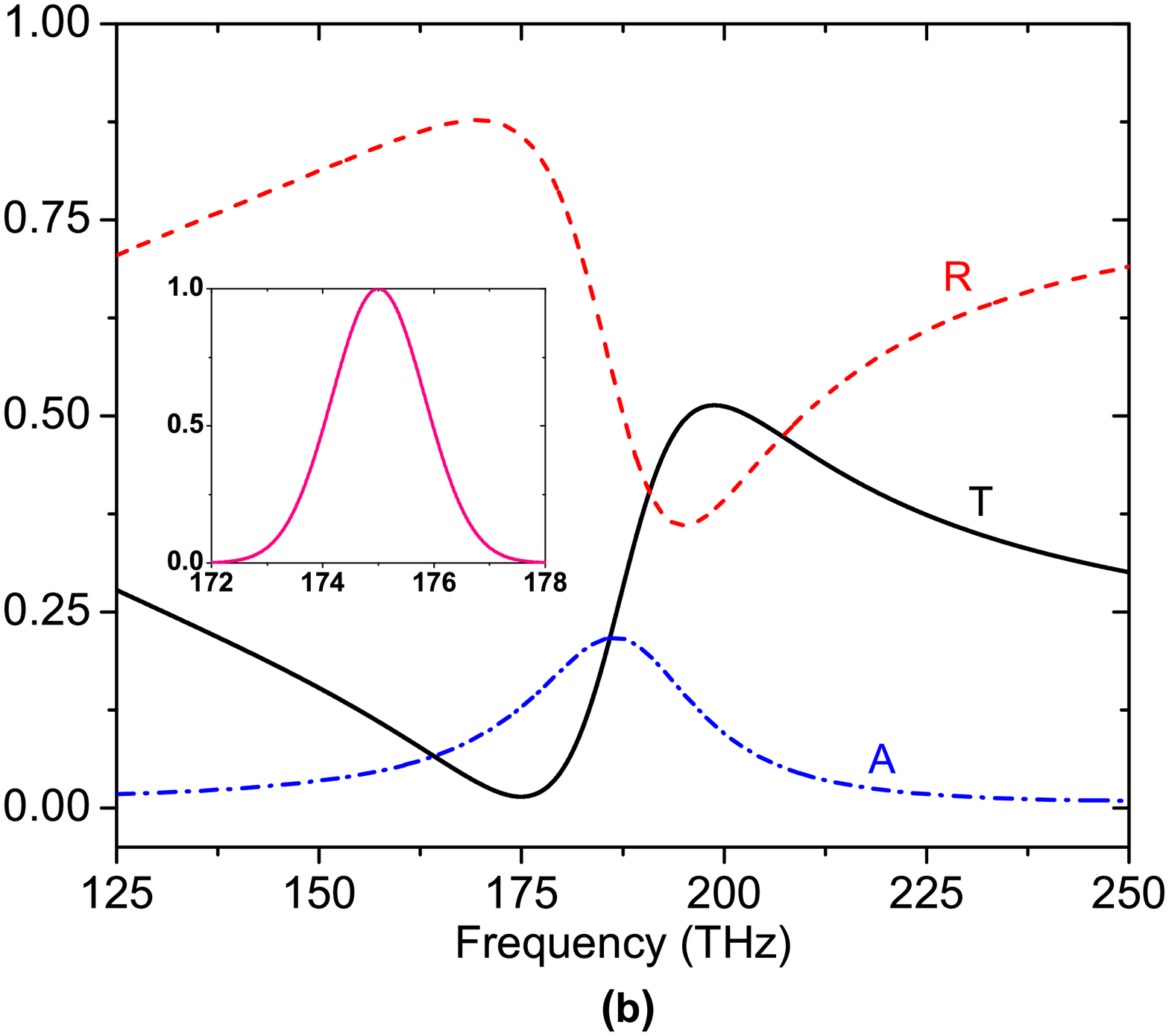}
 \caption{(Color online)
 (a) Schematic of the unit cell for the silver-based SRRs structure (yellow) with the
 electric field polarization parallel to the gap.
 The dielectric constants $\varepsilon$ for gain (red) and GaAs (light blue) are
 $9.0$ and $11.0$, respectively.
 (b) Calculated spectra for transmittance $T$ (black), reflectance $R$ (red), and
 absorptance $A$ (blue) for the structure shown in Fig.~\ref{fig2}(a).
 The insert shows the profile of the probe pulse with a center frequency
 of $175\mathrm{THz}$ ($\mathrm{FWHM}=2\mathrm{THz}$).
 }
 \label{fig2}
\end{figure}

The object of our studies is to present pump-probe simulations on arrays of silver
SRRs coupled to single quantum wells \cite{Meinzer2010,Meinzer2011}.
The structure considered is a U-shape SRRs fabricated on a gain-GaAs substrate
with a square periodicity of $p=250$nm (see Fig.~\ref{fig2}(a)).
The SRRs is made of silver with its permittivity modeled by a Drude responds:
$\epsilon(\omega)=1-\omega_{p}^{2}/(\omega^{2}+i\omega\gamma)$, with
$\omega_{p} = 1.37\times10^{16}$ rad/s and $\gamma = 2.73\times10^{13}$ rad/s.
The incident wave propagates perpendicular to the SRRs plane and has the
electric field polarization parallel to the gap (see Fig.~\ref{fig2}(a)).
The corresponding geometrical parameters are $a=150$nm, $h_{d}=40$nm, $h_{g}=20$nm,
$h_{s}=30$nm, $w=50$nm, and $h=75$nm.
Fig.~\ref{fig2}(b) shows the calculated spectrum (without pump) of  transmittance $T$,
reflectance $R$, and absorptance $A$ for the structure shown in Fig.~\ref{fig2}(a). The
resonant frequency is around $175\mathrm{THz}$, and we refer to the resonant frequency
according to the dip of the transmittance.
\begin{figure}[htbp]
 \centering
 \includegraphics[width=0.45\textwidth]{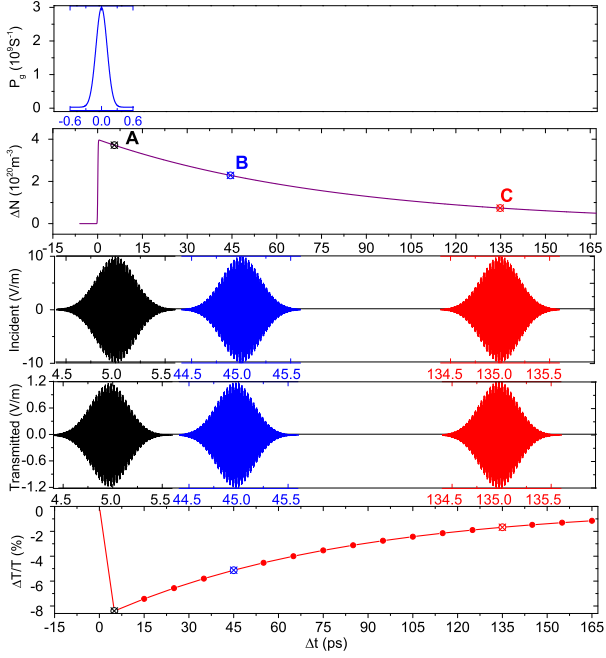}
 \caption{(Color online)
 Schematic of the numerical pump-probe experiments for the case on
 resonance. From the top to the bottom, each row is corresponding to the pump
 pulse, population inversion, incident signal (with time delays 5ps, 45ps and
 135ps), transmitted signal, and differential transmittance $\Delta T/T$. It
 should be mentioned here the incident frequency of the probe pulse is $175\mathrm{THz}$
 with $\mathrm{FWHM}$ of $2\mathrm{THz}$ and is equal to the SRRs resonance frequency.
 }
 \label{fig3}
\end{figure}
In our analysis, we first pump the active structure (see Fig.~\ref{fig2}(a))
with a short intensive Gaussian pump pulse, $P_g(t)$,  (see Fig.~\ref{fig3}, top panel).
After a suitable time delay (i.e. the pump-probe delay), we probe the structure with
a weak Gaussian probe pulse with a center frequency close to the SRRs resonance frequency
of $175\mathrm{THz}$.
Typical examples for the spatial distribution of electric field an gain are 
shown in \cite{SupplMat}.
The incident electric field amplitude of the probe pulse is $10\mathrm{V/m}$,
which is well inside the linear response regime.
Then, we can Fourier transform the time-dependent transmitted electric field and
divide by the Fourier transform of the incident probe pulse to obtain the spectral transmittance
of the system as seen by the probe pulse. Additionally, we obtain the total pulse transmittance
by dividing the energy in the transmitted pulse by the energy in the incident pulse,
integrated in time domain.
We define the differential transmittance, $\Delta T/T$, by taking the difference of
the measured total plus transmittance with pumping the active structure minus the same without pumping
and dividing it by the total plus transmittance without pumping.
This differential transmittance is a function of the pump-probe delay.
The bottom panel in Fig.~\ref{fig3} gives a differential transmittance $\Delta T/T$ which is negative.
this result was not expected, and we need to understand this behavior,
which agrees with the experiments \cite{Meinzer2010,Meinzer2011}.

\begin{figure}[htbp]
 \includegraphics[width=0.24\textwidth]{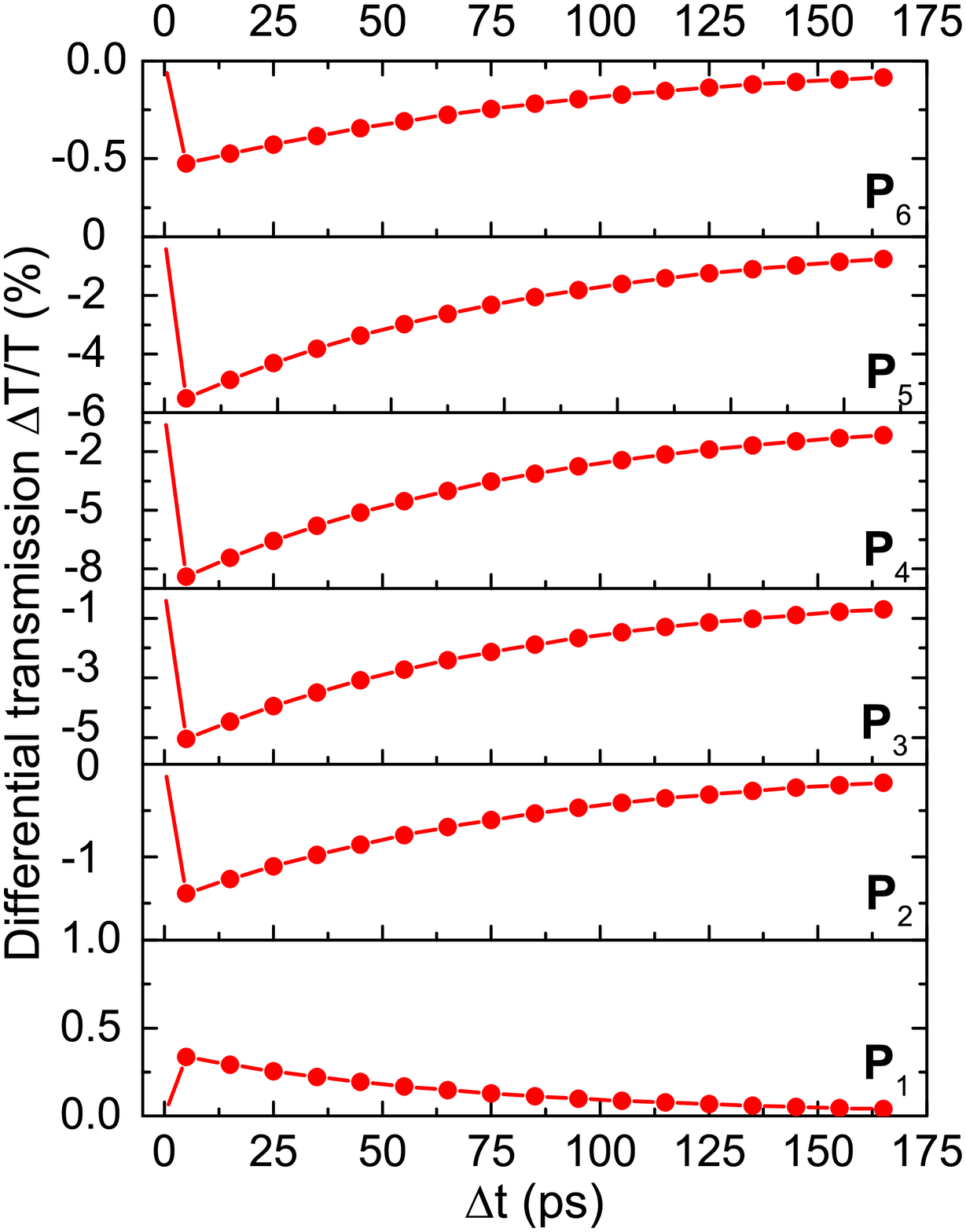}%
 \includegraphics[width=0.24\textwidth]{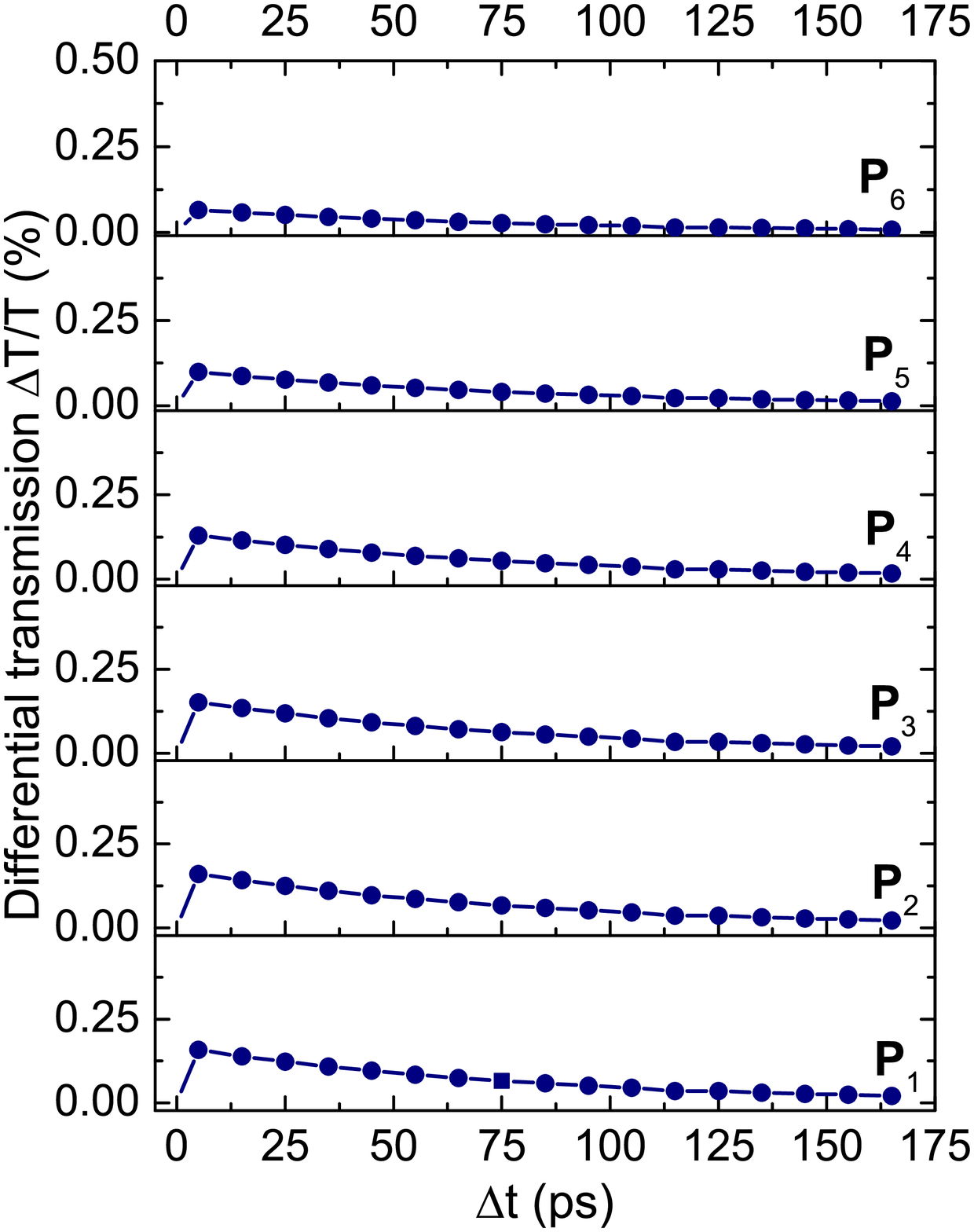}
 \caption{(Color online)
 Time domain numerical pump-probe experiments results for the SRRs that is nearly
 on-resonant with the gain material.  The left column corresponds to the
 parallel probe polarization with respect to the gap of the SRRs,
 the right column is the case for bare
 gain material, i.e., without SRRs on the top of the substrate. The width of the
 probe signal is $2\mathrm{THz}$ with decreasing in the probe center frequency from
 $179\mathrm{THz}$ for the top panel to $169\mathrm{THz}$ for the bottom panel.}
 \label{fig4}
\end{figure}

Fig.~\ref{fig4} gives an overview over the results obtained for the case of the SRRs on
resonance, i.e., $\omega_{a}=2\pi\times175\times10^{12}$ rad/s.
Data for the structure in Fig.~\ref{fig2}(a) (left column in Fig.~\ref{fig4})
and for the bare gain case (right column in Fig.~\ref{fig4}) without the SRRs on top
is shown.
For parallel polarization, the light does couple to the fundamental SRRs
resonance, for perpendicular polarization it does not.
The probe center frequency decreases from top (179THz) to bottom (169THz).
Note that the width of the
probe spectrum is $2$THz (see the insert of Fig.~\ref{fig2}(a)). Hence, the data have been
taken with $2$-THz spectral separation.  Inspection of the left column shows a
rather different behavior for the SRRs with gain compared to the bare gain case.
While the bare gain always delivers positive $\Delta T/T$ signals below $+
0.16\%$ (right column) over the whole probe spectrum.  The sign and magnitude
of the signals change for the case SRRs with gain. Under some conditions,
$\Delta T/T$ reaches values as negative as $- 8.50\%$ around
$f_{probe}=175$THz. Additionally, we may also get positive $\Delta T/T$ at the
very edges of the probe range (see left column in Fig.~\ref{fig4}).
If we turn to the
case of perpendicular polarization case, no distinct change between the
pump-probe results on the SRRs (not shown in Fig.~\ref{fig4}) and the bare gain
(right column in Fig.~\ref{fig4}), neither in the magnitude nor in the dynamics of the
$\Delta T/T$, can be detected.

\begin{figure}[htbp]
 \centering\includegraphics[width=0.45\textwidth]{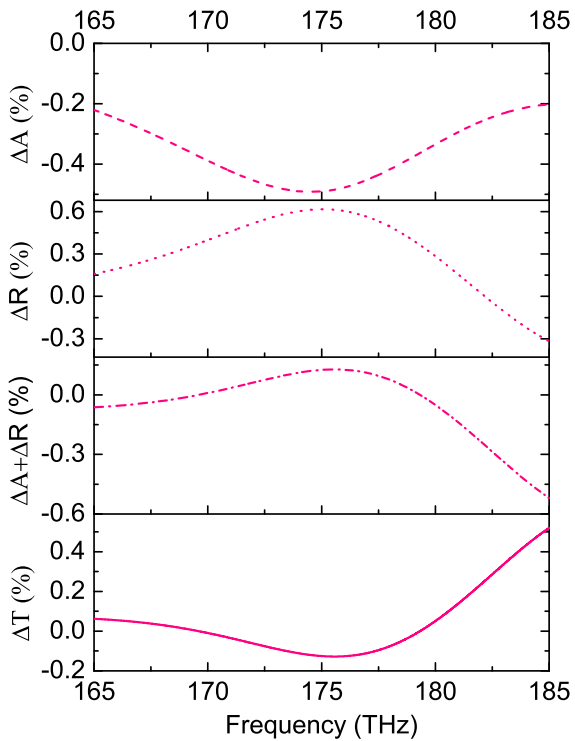}
 \caption{ (Color online) Frequency domain numerical pump-probe experiments results for the on-resonance case.
  Simulations results for the differences in transmittance ($\Delta T$), reflectance ($\Delta R$), and absorptance ($\Delta A$)
  versus frequency.
 }
 \label{fig5}
\end{figure}

We argue the distinct behavior can be attributed to the strong coupling between
the resonances of the SRRs and the gain medium. The negative $\Delta T/T$ are
not as we expected at first glance: the pump lifts electrons from ground state
to an excited state so that the absorption of the probe pulse is reduced,
leading to an increase of transmission. This is not the whole story. The
reason lies in the fact that with pump we not only affect the absorption, but
disturb the reflection of the structure, resulting in the mismatching of the
impedance.  Furthermore, we observed either increasing or decreasing tendency
for the case of on resonance as shown in Fig.~\ref{fig4}. All those behaviors can be
explained by the competing of the weak gain resonance and the impedance
mismatching between pump and without pump cases.  We will explore the
underlying mechanism below.
Fig.~\ref{fig5} shows the results for the difference in absorptance ($\Delta A$),
difference in reflectance ($\Delta R$), their sum ($\Delta A + \Delta R$), and
the difference in transmittance ($\Delta T = -(\Delta A + \Delta R)$) between
pump ($P_0=3\times{10}^{9} s^{-1}$) and no-pump using a wide probe ($\mathrm{FWHM}=54\mathrm{THz}$) pulse 
with a fixed pump-probe delay of $5\mathrm{ps}$.
As expected, we may observe a positive differential transmittance, $\Delta T/T>0$,
when we pump the gain, $\Delta A<0$, and if $\Delta R$ (impedance match) remains unchanged.

\begin{figure}[htbp]
 \centering\includegraphics[width=0.45\textwidth]{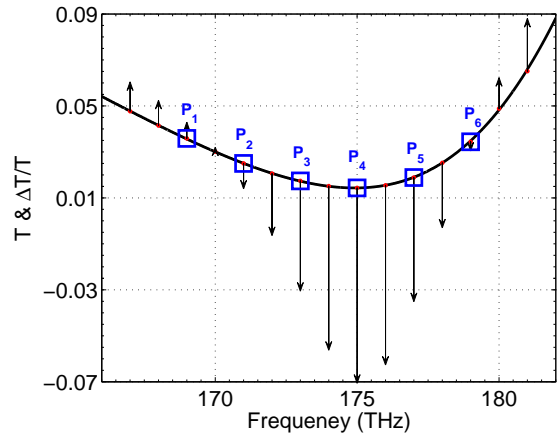}
 \caption{(Color online) The transmittance T (without pump, solid line) and the on-resonance
 differential transmittance $\Delta T/T$ results (vector arrow). The direction
 and the length of the arrow stand for the sign and the amplitude of $\Delta
 T/T$, respectively. The squares from $\textbf{P}_{1}$ to $\textbf{P}_{6}$
 correspond to the frequency of probe pulse ranging from $169\mathrm{THz}$ to $179\mathrm{THz}$ with
 uniform step of $2\mathrm{THz}$.
 }
 \label{fig6}
\end{figure}

The results of Fig.~\ref{fig5} are obtained for pump-probe experiments with the probe frequency equal to
the resonance frequency of the SRRs ($175\mathrm{THz}$) at a pump-probe delay of $5\mathrm{ps}$; 
Results for longer pump-probe delays are shown in supplementaty material \cite{SupplMat}.
Notice that $\Delta R$ is positive, $\Delta A$ is negative, and $\Delta T$ is also negative very close
the the resonance frequency.
If the probe center frequency moves away from the SRRs resonance frequency, the negative
$\Delta T/T$ decreases in magnitude and finally $\Delta T/T$ becomes positive.
These results are shown in Fig.~\ref{fig6} and agree with experiments \cite{Meinzer2010,Meinzer2011}.
If we can increase the magnitude of the Gaussian pump pulse $P_g(t)$ to $5\times{10}^{10} s^{-1}$ and
we repeat the pump-probe experiments, $\Delta T/T \approx -100\%$ at resonance frequency, $175\mathrm{THz}$.
If we increase the pump amplitude further to ${10}^{11} s^{-1}$ we can compensate the losses.
However, such pump intensities are unrealistic experimentally \cite{SupplMat}.
In conclusion, we have introduced a new approach for pump-probe simulations of metallic
metamaterials coupled to gain materials. We study the coupling between the U-shaped SRRs and the gain
material described by a 4-level gain model.
Using pump-probe simulations we find a distinct behavior for the differential transmittance $\Delta T/T$
of the probe pulse with and without SRRs both in magnitude and sign (negative, unexpected, and/or positive).
Our new approach has verified that the coupling between the metamaterial resonance and the gain medium
is dominated by near-field interactions.
Our model can be used to design new pump-probe experiments to compensate the losses of metamaterials.
\section*{Acknowledgements}
Work at Ames Laboratory was supported by the U. S. Department of Energy (Basic Energy Science, 
Division of Materials Sciences and Engineering) under contract No.~DE-ACD2-07CH11358.
This work was partially supported by the European Community FET project PHOME (No.~213390) and by
Laboratory-Directed Research and Development Program at Sandia National Laboratories.
The author Zhixiang Huang gratefully acknowledges the support of
the National Natural Science Foundation of China (No.~60931002, 61101064),
Distinguished Natural Science Foundation (No.~1108085J01) and
Universities Natural Science Foundation of Anhui Province (No.~KJ2011A002).

\bibliographystyle{apsrev}

\end{document}